\documentclass[aps, prb, twocolumn, showkeys]{revtex4-1}

\usepackage{chemformula} 
\usepackage{siunitx} 
\usepackage[inline,shortlabels]{enumitem} 
\usepackage{hyperref} 
\usepackage{graphicx} 
\usepackage{natbib}

\begin{document}

    \title{Restoration of long range order of \ch{Na} ions in \ch{Na_xCoO2} at high temperatures by sodium site doping}
    \author{M. H. N. Assadi}
    \email{assadi@aquarius.mp.es.osaka-u.ac.jp}
    \altaffiliation[Tel:]{+81668506671}
    \author{H. Katayama-Yoshida}
    \affiliation{Graduate School of Engineering Science, Osaka University, Osaka 560-8531, Japan}
    \date{2015}

    \begin{abstract}
        We have systematically investigated the \ch{Na_xCoO2} system doped with \ch{Cu}, \ch{Y}, \ch{Sn}, \ch{W}, \ch{Au} and \ch{Bi} for $x = 0.5, 0.75$ and 1.00 using density functional theory. Sn, W, and Bi always substitute a Co while Au always substitutes a Na regardless of Na concentration. However, for Cu and Y, the substitution site depends on Na concentration. When compared to the available experimental data, we find that thermoelectric performance is enhanced when the dopants substitute a Na site. In this case, surprisingly, resistivity decreases despite the reduced hole concentration caused by carrier recombination. We propose improved carrier mobility to be the cause of observed reduced resistivity.
    \end{abstract}
    \maketitle

    \section{Introduction}
\begin{figure}
            \centering
            \includegraphics[width=0.95\columnwidth]{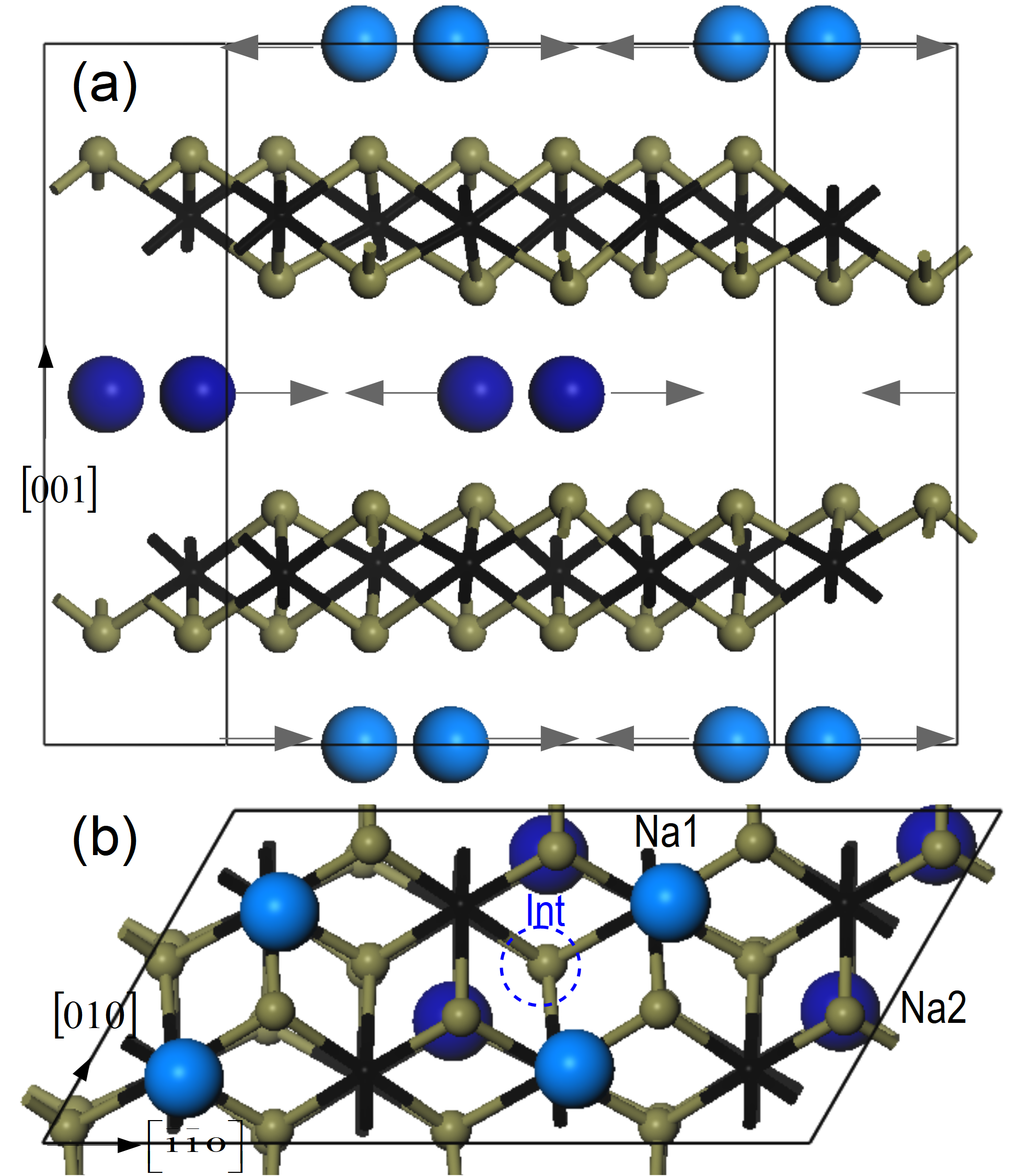}
            \caption{\label{fig:1}The top (a) and the side (b) view of the \ch{Na_{0.5}CoO2} supercell. The black, grey and blue spheres represent Co, O and Na ions. The direction of Na motion is marked with grey arrows.}
        \end{figure}

        Layered sodium cobaltate (\ch{Na_xCoO2}) is a fascinating material that has exotic magnetic, thermoelectric and superconducting properties.
        For $0.5 < x < 1$, polycrystalline \ch{Na_xCoO2} has a comparatively high figure of merit ($ZT$) of $\sim 0.8$ at temperature ($T$) range of $\sim 800-1000 \si{.K}$.
        This $ZT$ is unusually high for an oxide.\cite{Terasaki2013}
        As shown in Fig. \ref{fig:1}, \ch{Na_xCoO2} lattice consists of alternating Na layers and edge-sharing \ch{CoO6} octahedral layers.
        In Na deficient systems ($x < 1$), the Na ions adopt various configurations within their plane lowering crystal's symmetry.\cite{Wang2007}
        Furthermore, due to high Na diffusivity (schematically indicated by grey arrows in Fig. \ref{fig:1}), the Na layer is highly disordered by randomly distributed vacant sites in temperatures above ambient thus creating a medium that disturbs the propagation of phonon excitations.\cite{Koumoto2006, Roger2007} This irregular and dynamic positioning of the Na ions, therefore, considerably reduces the lattice thermal conductivity ($\kappa_L$) to $\sim 0.01 \si{.Wcm^{-1}K^{-1}}$ at $T =\,\sim\!1000 \si{.K}$.\cite{Nagira2003} This is in contrast to most oxides in which the dominance of covalent bonding causes relatively high $\kappa_L$; for instance, \ch{ZnO} has a $\kappa_L$ value of $\sim 1.25 \si{.Wcm^{-1}K^{-1}}$ at $T =\,\sim\!1000 \si{.K}$.\cite{Ozgur2006}
        However, the irregular pattern of Na ions comes with the disadvantage of low carrier mobility which is in the range of $\sim 0.8 \si{.cm^2 V^{-1}s^{-1}}$ for \ch{Na_xCoO2} at room temperature \cite{Brinks2014} (compared to $\sim 205 \si{.cm^2 V^{-1}s^{-1}}$ of \ch{ZnO} at the same temperature range).\cite{Look1998}
        The lower charge carrier mobility is nonetheless partly compensated by high carrier concentrations ($n$) which is in order of $\sim 10^{+21}$ to $10^{+22} \si{.cm^{-3}}$ for \ch{Na_{0.5}CoO2}.\cite{Terasaki2002}
        We should notice that this level of carrier concentration is also unusually high for a good thermoelectric semiconductor which has been demonstrated to be $\sim 10^{19}\si{.cm^{-3}}$.

\begin{figure}
            \centering
            \includegraphics[width=0.95\columnwidth]{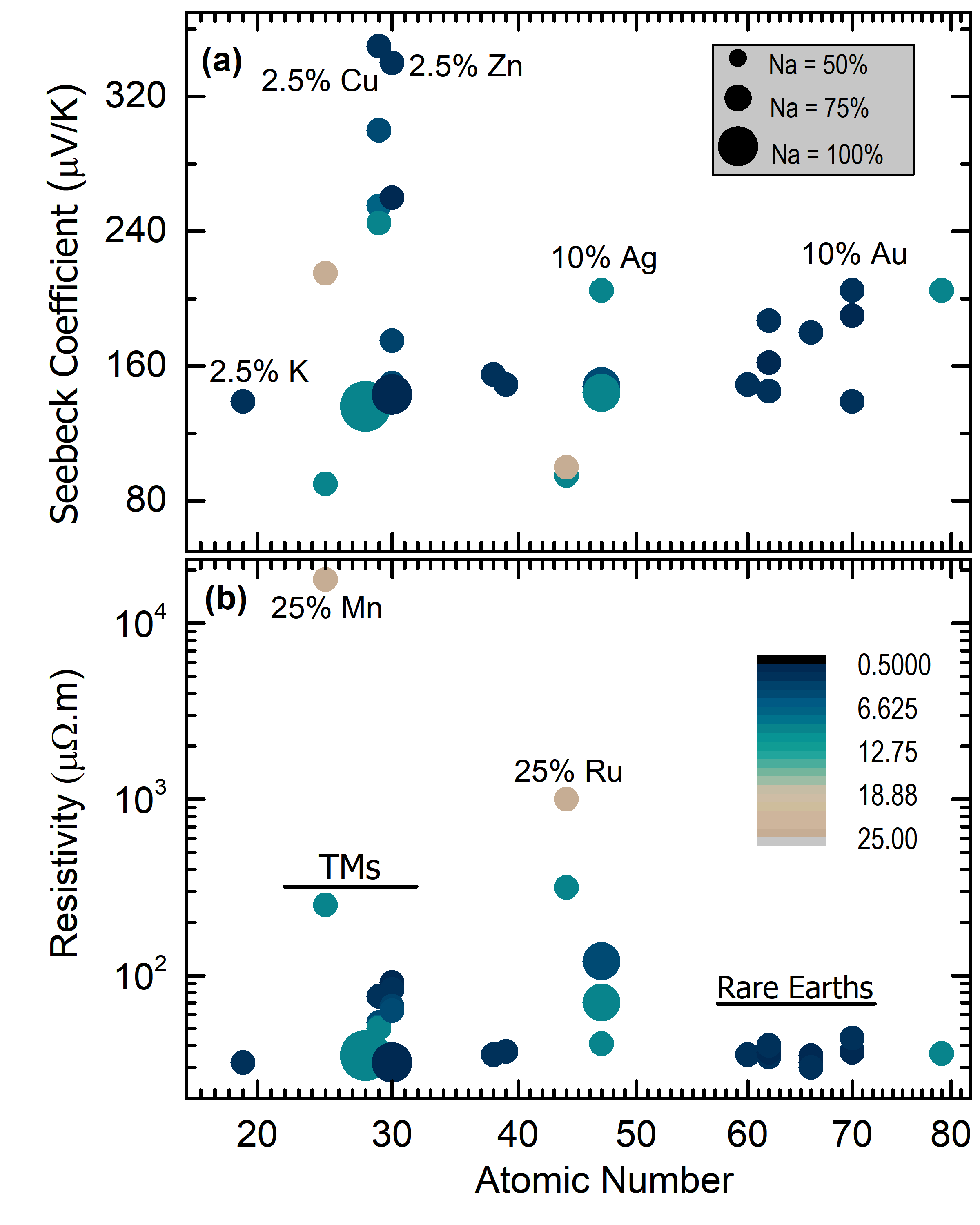}
            \caption{\label{fig:2}The resistivity and Seebeck coefficient of doped \ch{Na_xCoO2} as reported in the literature at $T = \,\sim \! 800\si{.K}$. The radius of the circles proportionally correlates to the Na concentration as demonstrated in the legend of (a). The shade of the circles represents the concentration of the dopants as expressed in the legend of (b). The data are taken from the following works: K,\cite{Nagira2003} Mn, \cite{Li2000} Ni, \cite{Wang2009} Cu, \cite{Park2006} Zn, \cite{Park2008, Tsai2011} Sr and Y, \cite{Nagira2003} Ru, \cite{Li2000} Ag, \cite{Seetawan2006, Ito2008} Nd, \cite{Nagira2003} Sm, \cite{Nagira2003, Nagira2004} Dy, \cite{Nagira2004} Yb, \cite{Nagira2003, Nagira2004} and Au. \cite{Ito2008}}
        \end{figure}

        In order to further improve the thermoelectric performance of \ch{Na_xCoO2}, doping has extensively been used to increase the $ZT$ of \ch{Na_xCoO2}.
        Nonetheless, the choice of dopants has usually been restricted by the solubility limits and mechanical and chemical restrictions imposed by the fabrication techniques.
        As a result, the experimental advancement in the doped \ch{Na_xCoO2} has been mainly guided by the practical consideration rather comprehensive and strategic attempt to improve the factors influencing the $ZT$.
        Consequently, not only the initial ambition of a figure of merit comfortably greater than one was not realized, but also many theoretically interesting questions remained unanswered too.
        One of these questions that we try to address here is how carrier mobility is affected by the dopants and how this is reflected on the thermoelectric performance.
        Fig. \ref{fig:2} shows the Seebeck coefficient (a) and electrical resistivity (b) of \ch{Na_xCoO2} doped with a wide variety of dopants as reported in the literature for $T =\,\sim\!800\si{.K}$.\cite{Nagira2003, Li2000, Nagira2004, Seetawan2006, Ito2006, Park2006, Ito2008, Park2008, Wang2009, Tsai2011}
        After inspection, we notice that
        \begin{enumerate*}[(i)] \item the highest Seebeck coefficient is obtained for late transition elements dopants like Cu and Zn for which the Seebeck coefficient exceeds $\,\sim\! 320  \si{.\mu V K^{-1}}$, followed by noble metals dopants such as Ag and Au for which the Seebeck coefficient is $\,\sim\! 200  \si{.\mu V K^{-1}}$ and the rare earth dopants for which the Seebeck coefficient is $\,\sim\! 180  \si{.\mu V K^{-1}}$;
        \item The highest Seebeck coefficient is achieved when $x =\,\sim\! 0.5$ as higher Na concentrations consistently result in lower Seebeck coefficient;
        \item the lowest resistivity is achieved for dopant concentrations of $\sim 2.5 \%$ or lower. Higher dopant concentrations result in the higher resistivity ($\rho$), sometimes by few orders of magnitude.
            To interpret these experimental results, we conducted a comprehensive theoretical study of the \ch{Na_xCoO2:M} system in which $x = 0.5, 0.75$ and 1.00 while M = Cu, Y, Sn, W, Au and Bi, representing different element groups.
        \end{enumerate*}

    \section{Computational details}
        We performed \textit{ab initio} spin-polarized density functional calculations using Accelrys's \ch{DMol^3} package.\cite{Delley1990, Delley2000}
        Self-consistent energy calculations were performed with double-numeric plus polarization basis and generalized gradient approximation based on Perdew-Wang formalism for the exchange-correlation functional.\cite{Perdew1992}
        Real-space global cut-off radii were set for all elements at $6.00 \si{\angstrom}$, and Brillouin zone sampling was carried out by choosing a $2\times4\times2$ $k$-point set within Monkhorst-Park scheme with a grid spacing of $\sim 0.05 \si{.\angstrom^{-1}}$ between $k$ points.
        The geometry optimization convergence thresholds for energy, Cartesian components of internal forces acting on the ions, and displacement were set to be $10^{-5} \si{.eV/atom}$, $0.01 \si{.eV/\angstrom}$, and $0.005 \si{.\angstrom}$.
        A $4a \times 2a \times 1c$ supercell of \ch{Na16Co16O32} constructed for studying the doped systems.
        To vary sodium concentration, four or eight sodium ions were removed from the original supercell to create \ch{Na12Co16Co32} and \ch{Na8Co16O32} supercells for which $x = 0.75$ and 0.50 respectively.
        The overall crystal structure of \ch{Na_xCoO2} and the ordering of Na ions critically depends on the Na concentration and has been extensively investigated both experimentally and theoretically.\cite{Zandbergen2004, Zhang2005, Meng2005}
        In this work, the Na patterns in the supercell for different values of $x$ were adopted after our previous work.\cite{Assadi2015}
        Since we only considered dopants that replaced cations, dopants' formation energy ($E^f$) was calculated for four possible geometric configurations.
        In the first configuration, M substituted Na ion at Na1 site creating a \ch{M_{Na 1}} configuration.
        In the second configuration, M substituted a Na ion at Na2 site creating a \ch{M_{Na 2}} configuration.
        Na1 shares basal with Co and occupies the Wyckoff site $b$ while Na2 site shares the basal coordinates with O and occupies the Wyckoff site $d$ of the $P6_3/mmc$ \ch{NaCoO2} primitive cell as demonstrated in Fig. \ref{fig:1}.
        In the third configuration, M occupied an interstitial site in Na layer creating a \ch{M_{Int}} configuration.
        Few distinct positions for the \ch{M_{Int}} configuration were considered.
        We found that the interstitial dopants are more stable when located on a vacant Na2 (or Wyckoff $d$) position.
        Finally, the fourth configuration is constructed by M substituting a Co ion creating a \ch{M_{Co}} configuration.
        One special case was the \ch{M_{Na 1}} for $x = 1$; since in \ch{NaCoO2}, all Na ions were located in Na2 location, no stable local minimum for \ch{M_{Na 1}} was found; all \ch{M_{Na 1}} dopants instead relaxed to Na2 location.
        The formation energy ($E^f$) was calculated using the standard procedure as described by the following equation:
        \begin{equation}
            \label{eq:formation_energy}
            E^f = E^t \left(\ch{Na_xCoO2:M}\right) + \mu_\alpha - E^t \left(\ch{Na_xCoO2}\right) - \mu_M
        \end{equation}
        Here, $E^t \left(\ch{Na_xCoO2:M}\right)$ is the total energy of the \ch{Na_xCoO2} supercell containing the dopant M and $E^t \left(\ch{Na_xCoO2}\right)$ is the total energy of the pristine \ch{Na_xCoO2} supercell.
        $\mu_\alpha$ and $\mu_M$ are the chemical potentials of the removed and added elements respectively.
        The chemical potentials were calculated from the total energies of their most stable oxides representing an oxygen-rich condition except for Au were the chemical potential was calculated from the total energy of Au's metallic form. The formation energy of the dopants in \ch{Na_xCoO2} is presented in Fig. \ref{fig:3} (a)-(f).

    \section{Results and discussion}
        In the case of Cu doping, for $x = 0.5$, \ch{Cu_{Int}} had the lowest formation energy of 1.45 eV.
        However, for $x = 0.75$ and 1.00, the most stable configuration was \ch{Cu_{Co}} with formation energy of 2.49 eV and 1.29 eV respectively.
        In the \ch{Na_xCoO2:Y} system, \ch{Y_{Na 1}} had the lowest formation energy of 2.59 eV for $x = 0.5$.
        For higher Na concentrations, on the other hand, in a trend similar to the case of Cu doping, the most stable configuration was \ch{Y_{Co}} with formation energy of 3.94 eV for $x = 0.75$ and 3.74 eV for $x = 1.00$.
        In the case of Sn doping, \ch{Sn_{Co}} always had the lowest formation energy for all considered Na concentrations.
        In this case, \ch{Sn_{Co}} had an $E^f$ of $-1.06\si{.eV}, -0.94\si{.eV}$ and $-0.53\si{.eV}$ for $x = 0.5, 0.75$ and 1.00 respectively.
        For the \ch{Na_xCoO2:W} system, \ch{W_{Co}} was the most stable configuration for all values of $x$ having an $E^f$ of 3.68 eV, 4.40 eV and 4.86 eV for $x = 0.50, 0.75$ and 1.00 respectively.
        In the \ch{Na_xCoO2:Au} system, \ch{Au_{Int}} was the most stable structure for $x = 0.5$ with an $E^f$ of 0.87 eV.
        For $x = 0.75$ the most stable configuration was \ch{Au_{Na 1}} with an $E^f$ of 1.88 eV and for $x = 1.00$, \ch{Au_{Na 2}} was the most stable configuration with an $E^f$ of 2.01 eV.
        For the \ch{Na_xCoO2:Bi} system, \ch{Bi_{Co}} was the most stable configuration for all considered Na concentration having an $E^f$ of 2.16 eV, 2.89 eV and 3.89 eV for $x = 0.5, 0.75$ and 1.00 respectively.

    \begin{figure}
        \centering
        \includegraphics[width=0.95\columnwidth]{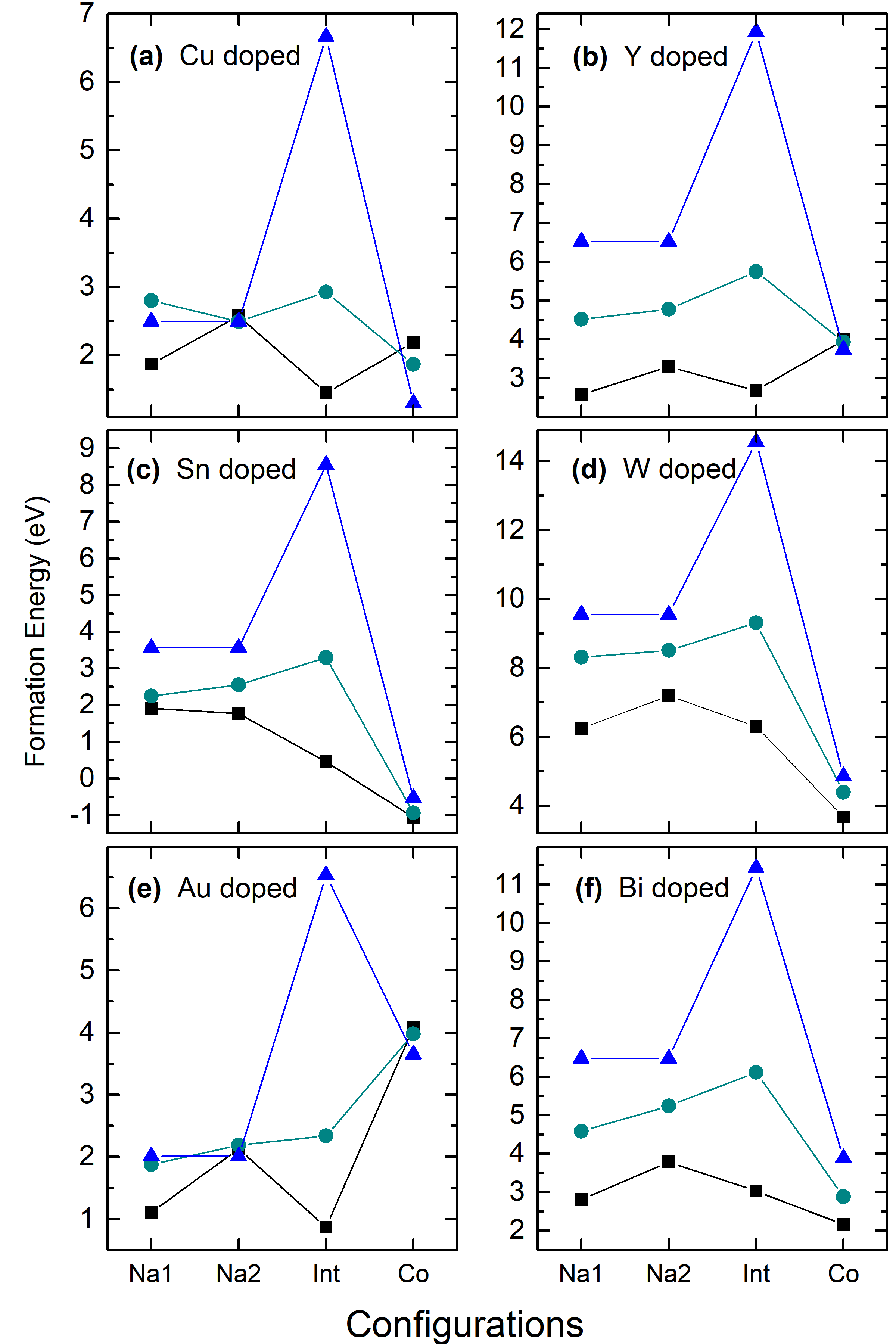}
        \caption{\label{fig:3}The formation energy ($E^f$) of dopants in \ch{Na_xCoO2}  host matrix for any given configuration,\textit{et al}. arranged according to the dopants' atomic number. The black, green and blue symbols denote $x = 0.50, 0.75$ and 1.00. The lines were drawn for visual guidance.}
    \end{figure}

        We can classify these dopants in three different categories; the ones that always substitute a Co regardless of Na concentration like Bi, W and Sn; the one that is always located in the Na layer namely Au regardless of Na concentration; and the ones for which the most stable configuration depends on Na concentration.
        The latter ones, Cu and Y, are located in the Na layer for lower Na concentration, i.e.\ $x = 0.5$, but they become more stable when substituting Co at higher Na concentrations.
        By comparing the DFT results with the experiments, we find that Cu and Au dopants that best improve the thermoelectric properties reside in the Na layer for $x = 0.5$.
        Furthermore, our previous investigation has also shown that the rare earth element Eu that improves the thermoelectric performance also resides on the Na layer.\cite{Assadi2013}

    \begin{figure}
        \centering
        \includegraphics[width=0.95\columnwidth]{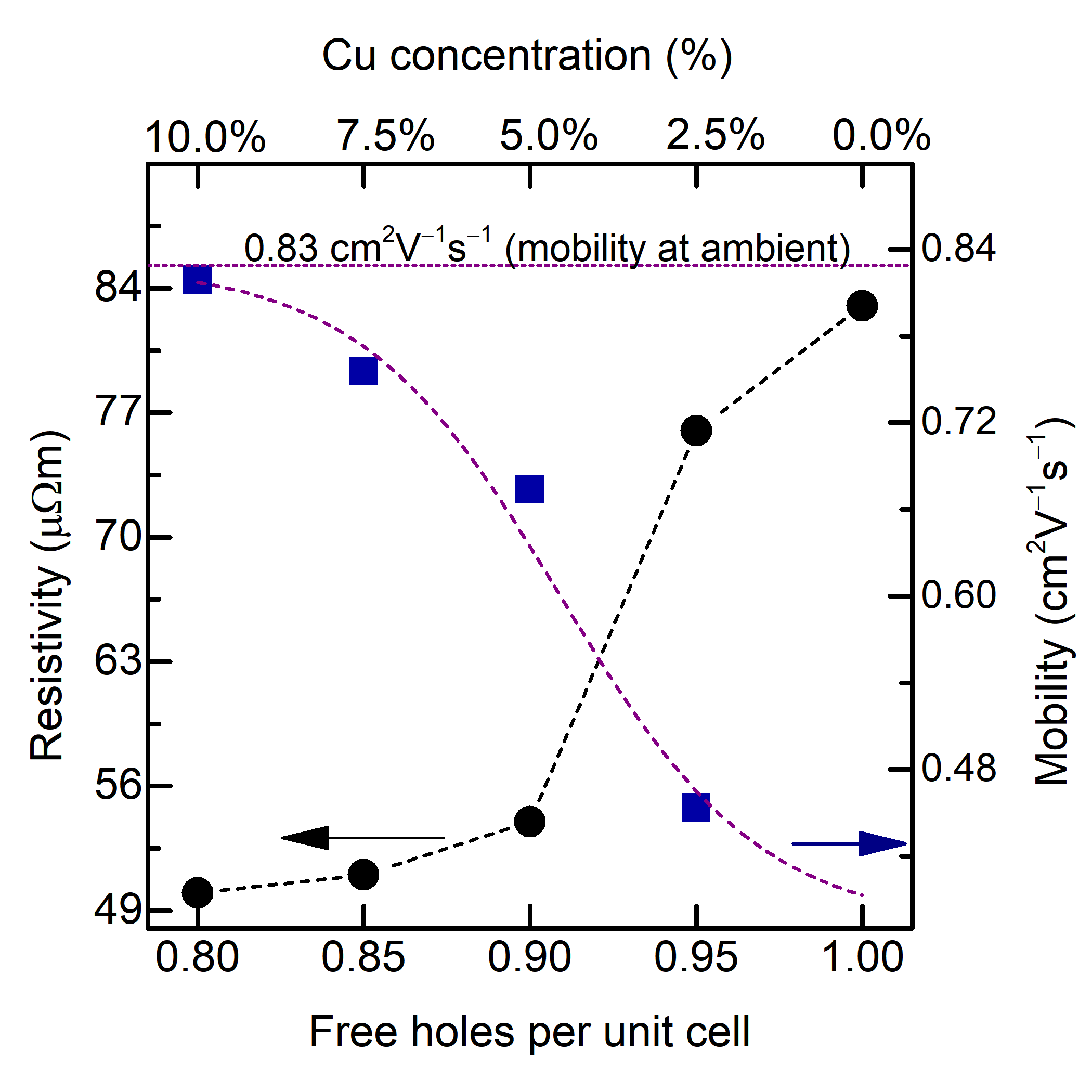}
        \caption{\label{fig:4}The black circles represent the resistivity measurement of \ch{Na_{0.5}CoO2:Cu} adopted from the work of Park \textit{et al}. \cite{Park2006} The hole density was calculated assuming that Cu is incorporated in the Ns layer interstitially. The mobility was calculated by assuming that only holes contribute to conduction. The mobility data were fitted with a growth function, and it was found that \ch{Cu_{Int}} raises the mobility to the values at ambient as measured by Brinks \textit{et al}.\cite{Brinks2014}}
    \end{figure}

        A dopant with an oxidation state more than 1+ that replaces a Na ion reduces the carrier concentration through electron-hole recombination by compensating for the Na vacancy in the \ch{Na_xCoO2} system (stoichiometric \ch{NaCoO2} is a band insulator).\cite{Lang2005}
        This is also true for any cationic dopant that is placed in Na layer interstitially.
        As a consequence, one anticipates an increase in the electrical resistivity by doping such an element.
        The experimental data, however, obviously indicate the opposite.
        In one instance, where comprehensive experimental data were available, the resistivity of undoped \ch{Na_{0.5}CoO2} was $83 \si{.\micro \Omega \cdot m} $ at $T = 800\si{.K}$ while introducing $2.5\%, 5\%, 7.5\%$ and $10\%$ of Cu reduced $\rho$ to $76 \si{.\micro \Omega \cdot m}, 54 \si{.\micro \Omega \cdot m}, 51 \si{.\micro \Omega \cdot m}$ and $50\si{.\micro \Omega \cdot m}$ respectively \cite{Park2006} as represented by black circles in Fig. \ref{fig:4}.
        A similar effect was also observed for Ag doped \ch{Na_{0.75}CoO2} where the $\rho$ decreased from $130 \si{.\micro \Omega \cdot m}$ for the undoped sample to $110\si{.\micro \Omega \cdot m}$ and $90 \si{.\micro \Omega \cdot m}$ for the \ch{Na_{0.75}CoO2} sample doped with $5\%$ and $10\%$ of Ag respectively.\cite{Seetawan2006}
        One plausible explanation for decreasing resistivity, in this case, is an increase in carrier's mobility.
        As previously reported, \ch{^{23}Na} NMR spectra and relaxation measurements suggest that for $T \geq 292\si{.K}$ the Na layers in undoped \ch{Na_xCoO2} adopt a 2D-liquid state, in which the Na ions lose the static structure with a well-defined allocation of sites.\cite{Weller2009}
        As a consequence, Na layers act as a strong scatterer for the carriers and result in poor carrier mobility.
        However, the incorporation of a heavier and more positively charged dopants in the Na layer creates both mass and electrostatic inertia against the highly mobile Na ions at higher temperatures.
        It seems that in the doped Cu doped system, the incorporation of the Cu dopants in the Na layer partially restores the long-range order among Na ions.
        Consequently, the mean free path of the charge carriers increases thus improving carriers' mobility.
        The restoration of long-range ordering has been directly verified for other dopants by experiments where the dopants are incorporated in the Na layer.
        For example, Raman spectroscopy measurement detected the peaks of Na ordering at room temperatures in \ch{Mg_{Na}} doped \ch{Na_{0.8}CoO2} system.
        These peaks were absent in the undoped samples at the same temperature.\cite{Tsai2012}
        Furthermore, neutron diffraction experiments have shown that Ca doping in Na layer creates a Na superlattice ordered over long range at temperatures as high as $490\si{.K}$.\cite{Porter2014}
        This in contrast to the behaviour of Na ions in undoped \ch{Na_{0.5}CoO2} in which the superstructures of tri-vacancy in Na layer vanishes at temperatures higher than 200 K. \cite{Voneshen2013}
        To further examine this idea, we calculated the carrier mobility for different Cu concentration in \ch{Na_{0.5}CoO2:Cu} at 800 K. \cite{Park2006}
        By examining the Mulliken charge population of \ch{Cu_{Int}} (the most stable configuration of Cu in \ch{Na_{0.5}CoO2} according to Fig. \ref{fig:3}(a)), we found that Cu's $3d$ orbitals were occupied by $\sim9.75$ electrons implying that Cu had an oxidation state of $\sim1+$.
        Then, accordingly, we adjusted the number of free carrier per unit cell and calculated the carrier concentration ($n$) using the experimental lattice parameters of \ch{Na_{0.5}CoO2}. \cite{Chen2004}
        We further assumed that conductivity is mainly due to the hole carriers and neglect the Na ionic contribution to the conductivity.
        This assumption is justified by the fact that the ionic conductivity in \ch{Na_xCoO2} is in order of $\sim 6 \si{.mS/cm}$ \cite{Mo2014} (equivalent of a resistivity of $\sim 1.67 \si{.\Omega\cdot m}$) which is $\sim 5$ orders of magnitude smaller than the electronic conductivity.
        Finally, we used the resistivity-mobility relationship $\rho =1/e\mu n$ to obtain the mobility presented by blue squares in Fig. \ref{fig:4}.
        We see that despite decreasing hole concentration, carrier mobility increases with rising Cu concentration approaching the values of pristine \ch{Na_xCoO2} at ambient. \cite{Brinks2014}
        Hence, we can see the direct correlation between the Cu dopant concentration and increased mobility.
        This interesting phenomenon nonetheless needs to be further investigated by direct experimental probes as it is important not only to the thermoelectric applications but to the rechargeable Na ion batteries.\cite{Han2015}

    \section{Conclusions}
        In conclusion, guided by experimental results, we performed DFT calculation for \ch{Na_xCoO2} system doped with Cu, Y, Sn, W, Au and Bi.
        We found that the best thermoelectric performance is achieved when the dopant is located in the Na layer within the \ch{Na_xCoO2} structure.
        Although these dopants reduce the hole concentration, they have a greater impact increasing carrier mobility, therefore, improving electrical conductivity which is an important contributing factor toward $ZT$.
        However, the effect of this category of dopants on thermal conductivity should be experimentally investigated to find the extent of the trade-off between improving carrier mobility and increasing lattice thermal conductivity.

    \begin{acknowledgements}
        This work was supported by the Japanese society for the promotion of Science. Computational resources were provided by Intersect Australia Ltd.
    \end{acknowledgements}

    \bibliography{refs}{}

\begin{thebibliography}{34}%
\makeatletter
\providecommand \@ifxundefined [1]{%
 \@ifx{#1\undefined}
}%
\providecommand \@ifnum [1]{%
 \ifnum #1\expandafter \@firstoftwo
 \else \expandafter \@secondoftwo
 \fi
}%
\providecommand \@ifx [1]{%
 \ifx #1\expandafter \@firstoftwo
 \else \expandafter \@secondoftwo
 \fi
}%
\providecommand \natexlab [1]{#1}%
\providecommand \enquote  [1]{``#1''}%
\providecommand \bibnamefont  [1]{#1}%
\providecommand \bibfnamefont [1]{#1}%
\providecommand \citenamefont [1]{#1}%
\providecommand \href@noop [0]{\@secondoftwo}%
\providecommand \href [0]{\begingroup \@sanitize@url \@href}%
\providecommand \@href[1]{\@@startlink{#1}\@@href}%
\providecommand \@@href[1]{\endgroup#1\@@endlink}%
\providecommand \@sanitize@url [0]{\catcode `\\12\catcode `\$12\catcode
  `\&12\catcode `\#12\catcode `\^12\catcode `\_12\catcode `\%12\relax}%
\providecommand \@@startlink[1]{}%
\providecommand \@@endlink[0]{}%
\providecommand \url  [0]{\begingroup\@sanitize@url \@url }%
\providecommand \@url [1]{\endgroup\@href {#1}{\urlprefix }}%
\providecommand \urlprefix  [0]{URL }%
\providecommand \Eprint [0]{\href }%
\providecommand \doibase [0]{http://dx.doi.org/}%
\providecommand \selectlanguage [0]{\@gobble}%
\providecommand \bibinfo  [0]{\@secondoftwo}%
\providecommand \bibfield  [0]{\@secondoftwo}%
\providecommand \translation [1]{[#1]}%
\providecommand \BibitemOpen [0]{}%
\providecommand \bibitemStop [0]{}%
\providecommand \bibitemNoStop [0]{.\EOS\space}%
\providecommand \EOS [0]{\spacefactor3000\relax}%
\providecommand \BibitemShut  [1]{\csname bibitem#1\endcsname}%
\let\auto@bib@innerbib\@empty
\bibitem [{\citenamefont {Terasaki}(2013)}]{Terasaki2013}%
  \BibitemOpen
  \bibfield  {author} {\bibinfo {author} {\bibfnamefont {I.}~\bibnamefont
  {Terasaki}},\ }\enquote {\bibinfo {title} {Layered cobalt oxides: correlated
  electrons for thermoelectrics},}\ in\ \href {\doibase
  10.1007/978-3-642-37537-8_3} {\emph {\bibinfo {booktitle} {Thermoelectric
  Nanomaterials}}},\ \bibinfo {series} {Springer Series in Materials Science},
  Vol.\ \bibinfo {volume} {182},\ \bibinfo {editor} {edited by\ \bibinfo
  {editor} {\bibfnamefont {K.}~\bibnamefont {Koumoto}}\ and\ \bibinfo {editor}
  {\bibfnamefont {T.}~\bibnamefont {Mori}}}\ (\bibinfo  {publisher} {Springer
  Berlin Heidelberg},\ \bibinfo {year} {2013})\ \bibinfo {type} {Book
  section}~\bibinfo {chapter} {3}, pp.\ \bibinfo {pages} {51--70}\BibitemShut
  {NoStop}%
\bibitem [{\citenamefont {Wang}\ and\ \citenamefont {Ni}(2007)}]{Wang2007}%
  \BibitemOpen
  \bibfield  {author} {\bibinfo {author} {\bibfnamefont {Y.}~\bibnamefont
  {Wang}}\ and\ \bibinfo {author} {\bibfnamefont {J.}~\bibnamefont {Ni}},\
  }\href@noop {} {\bibfield  {journal} {\bibinfo  {journal} {Phys. Rev. B}\
  }\textbf {\bibinfo {volume} {76}},\ \bibinfo {pages} {094101} (\bibinfo
  {year} {2007})}\BibitemShut {NoStop}%
\bibitem [{\citenamefont {Koumoto}\ \emph {et~al.}(2006)\citenamefont
  {Koumoto}, \citenamefont {Terasaki},\ and\ \citenamefont
  {Funahashi}}]{Koumoto2006}%
  \BibitemOpen
  \bibfield  {author} {\bibinfo {author} {\bibfnamefont {K.}~\bibnamefont
  {Koumoto}}, \bibinfo {author} {\bibfnamefont {I.}~\bibnamefont {Terasaki}}, \
  and\ \bibinfo {author} {\bibfnamefont {R.}~\bibnamefont {Funahashi}},\
  }\href@noop {} {\bibfield  {journal} {\bibinfo  {journal} {MRS Bull.}\
  }\textbf {\bibinfo {volume} {31}},\ \bibinfo {pages} {206} (\bibinfo {year}
  {2006})}\BibitemShut {NoStop}%
\bibitem [{\citenamefont {Roger}\ \emph {et~al.}(2007)\citenamefont {Roger},
  \citenamefont {Morris}, \citenamefont {Tennant}, \citenamefont {Gutmann},
  \citenamefont {Goff}, \citenamefont {Hoffmann}, \citenamefont {Feyerherm},
  \citenamefont {Dudzik}, \citenamefont {Prabhakaran}, \citenamefont
  {Boothroyd}, \citenamefont {Shannon}, \citenamefont {Lake},\ and\
  \citenamefont {Deen}}]{Roger2007}%
  \BibitemOpen
  \bibfield  {author} {\bibinfo {author} {\bibfnamefont {M.}~\bibnamefont
  {Roger}}, \bibinfo {author} {\bibfnamefont {D.~J.~P.}\ \bibnamefont
  {Morris}}, \bibinfo {author} {\bibfnamefont {D.~A.}\ \bibnamefont {Tennant}},
  \bibinfo {author} {\bibfnamefont {M.~J.}\ \bibnamefont {Gutmann}}, \bibinfo
  {author} {\bibfnamefont {J.~P.}\ \bibnamefont {Goff}}, \bibinfo {author}
  {\bibfnamefont {J.~U.}\ \bibnamefont {Hoffmann}}, \bibinfo {author}
  {\bibfnamefont {R.}~\bibnamefont {Feyerherm}}, \bibinfo {author}
  {\bibfnamefont {E.}~\bibnamefont {Dudzik}}, \bibinfo {author} {\bibfnamefont
  {D.}~\bibnamefont {Prabhakaran}}, \bibinfo {author} {\bibfnamefont {A.~T.}\
  \bibnamefont {Boothroyd}}, \bibinfo {author} {\bibfnamefont {N.}~\bibnamefont
  {Shannon}}, \bibinfo {author} {\bibfnamefont {B.}~\bibnamefont {Lake}}, \
  and\ \bibinfo {author} {\bibfnamefont {P.~P.}\ \bibnamefont {Deen}},\ }\href
  {\doibase 10.1038/nature05531} {\bibfield  {journal} {\bibinfo  {journal}
  {Nature}\ }\textbf {\bibinfo {volume} {445}},\ \bibinfo {pages} {631}
  (\bibinfo {year} {2007})}\BibitemShut {NoStop}%
\bibitem [{\citenamefont {Nagira}\ \emph {et~al.}(2003)\citenamefont {Nagira},
  \citenamefont {Ito}, \citenamefont {Katsuyama}, \citenamefont {Majima},\ and\
  \citenamefont {Nagai}}]{Nagira2003}%
  \BibitemOpen
  \bibfield  {author} {\bibinfo {author} {\bibfnamefont {T.}~\bibnamefont
  {Nagira}}, \bibinfo {author} {\bibfnamefont {M.}~\bibnamefont {Ito}},
  \bibinfo {author} {\bibfnamefont {S.}~\bibnamefont {Katsuyama}}, \bibinfo
  {author} {\bibfnamefont {K.}~\bibnamefont {Majima}}, \ and\ \bibinfo {author}
  {\bibfnamefont {H.}~\bibnamefont {Nagai}},\ }\href {\doibase
  10.1016/s0925-8388(02)00799-5} {\bibfield  {journal} {\bibinfo  {journal} {J.
  Alloy. Compd.}\ }\textbf {\bibinfo {volume} {348}},\ \bibinfo {pages} {263}
  (\bibinfo {year} {2003})}\BibitemShut {NoStop}%
\bibitem [{\citenamefont {Ozgur}\ \emph {et~al.}(2006)\citenamefont {Ozgur},
  \citenamefont {Gu}, \citenamefont {Chevtchenko}, \citenamefont {Spradlin},
  \citenamefont {Cho}, \citenamefont {Morkoc}, \citenamefont {Pollak},
  \citenamefont {Everitt}, \citenamefont {Nemeth},\ and\ \citenamefont
  {Nause}}]{Ozgur2006}%
  \BibitemOpen
  \bibfield  {author} {\bibinfo {author} {\bibfnamefont {U.}~\bibnamefont
  {Ozgur}}, \bibinfo {author} {\bibfnamefont {X.}~\bibnamefont {Gu}}, \bibinfo
  {author} {\bibfnamefont {S.}~\bibnamefont {Chevtchenko}}, \bibinfo {author}
  {\bibfnamefont {J.}~\bibnamefont {Spradlin}}, \bibinfo {author}
  {\bibfnamefont {S.~J.}\ \bibnamefont {Cho}}, \bibinfo {author} {\bibfnamefont
  {H.}~\bibnamefont {Morkoc}}, \bibinfo {author} {\bibfnamefont {F.~H.}\
  \bibnamefont {Pollak}}, \bibinfo {author} {\bibfnamefont {H.~O.}\
  \bibnamefont {Everitt}}, \bibinfo {author} {\bibfnamefont {B.}~\bibnamefont
  {Nemeth}}, \ and\ \bibinfo {author} {\bibfnamefont {J.~E.}\ \bibnamefont
  {Nause}},\ }\href {\doibase 10.1007/s11664-006-0098-9} {\bibfield  {journal}
  {\bibinfo  {journal} {J. Electron. Mater.}\ }\textbf {\bibinfo {volume}
  {35}},\ \bibinfo {pages} {550} (\bibinfo {year} {2006})}\BibitemShut
  {NoStop}%
\bibitem [{\citenamefont {Brinks}\ \emph {et~al.}(2014)\citenamefont {Brinks},
  \citenamefont {Rijnders},\ and\ \citenamefont {Huijben}}]{Brinks2014}%
  \BibitemOpen
  \bibfield  {author} {\bibinfo {author} {\bibfnamefont {P.}~\bibnamefont
  {Brinks}}, \bibinfo {author} {\bibfnamefont {G.}~\bibnamefont {Rijnders}}, \
  and\ \bibinfo {author} {\bibfnamefont {M.}~\bibnamefont {Huijben}},\ }\href
  {\doibase doi:http://dx.doi.org/10.1063/1.4901447} {\bibfield  {journal}
  {\bibinfo  {journal} {Appl. Phys. Lett.}\ }\textbf {\bibinfo {volume}
  {105}},\ \bibinfo {pages} {193902} (\bibinfo {year} {2014})}\BibitemShut
  {NoStop}%
\bibitem [{\citenamefont {Look}\ \emph {et~al.}(1998)\citenamefont {Look},
  \citenamefont {Reynolds}, \citenamefont {Sizelove}, \citenamefont {Jones},
  \citenamefont {Litton}, \citenamefont {Cantwell},\ and\ \citenamefont
  {Harsch}}]{Look1998}%
  \BibitemOpen
  \bibfield  {author} {\bibinfo {author} {\bibfnamefont {D.~C.}\ \bibnamefont
  {Look}}, \bibinfo {author} {\bibfnamefont {D.~C.}\ \bibnamefont {Reynolds}},
  \bibinfo {author} {\bibfnamefont {J.~R.}\ \bibnamefont {Sizelove}}, \bibinfo
  {author} {\bibfnamefont {R.~L.}\ \bibnamefont {Jones}}, \bibinfo {author}
  {\bibfnamefont {C.~W.}\ \bibnamefont {Litton}}, \bibinfo {author}
  {\bibfnamefont {G.}~\bibnamefont {Cantwell}}, \ and\ \bibinfo {author}
  {\bibfnamefont {W.~C.}\ \bibnamefont {Harsch}},\ }\href {\doibase
  http://dx.doi.org/10.1016/S0038-1098(97)10145-4} {\bibfield  {journal}
  {\bibinfo  {journal} {Solid State Commun.}\ }\textbf {\bibinfo {volume}
  {105}},\ \bibinfo {pages} {399} (\bibinfo {year} {1998})}\BibitemShut
  {NoStop}%
\bibitem [{\citenamefont {Terasaki}\ \emph {et~al.}(2002)\citenamefont
  {Terasaki}, \citenamefont {Tsukada},\ and\ \citenamefont
  {Iguchi}}]{Terasaki2002}%
  \BibitemOpen
  \bibfield  {author} {\bibinfo {author} {\bibfnamefont {I.}~\bibnamefont
  {Terasaki}}, \bibinfo {author} {\bibfnamefont {I.}~\bibnamefont {Tsukada}}, \
  and\ \bibinfo {author} {\bibfnamefont {Y.}~\bibnamefont {Iguchi}},\
  }\href@noop {} {\bibfield  {journal} {\bibinfo  {journal} {Phys. Rev. B}\
  }\textbf {\bibinfo {volume} {65}},\ \bibinfo {pages} {195106} (\bibinfo
  {year} {2002})}\BibitemShut {NoStop}%
\bibitem [{\citenamefont {Li}\ \emph {et~al.}(2000)\citenamefont {Li},
  \citenamefont {Funahashi}, \citenamefont {Matsubara},\ and\ \citenamefont
  {Sodeoka}}]{Li2000}%
  \BibitemOpen
  \bibfield  {author} {\bibinfo {author} {\bibfnamefont {S.~W.}\ \bibnamefont
  {Li}}, \bibinfo {author} {\bibfnamefont {R.}~\bibnamefont {Funahashi}},
  \bibinfo {author} {\bibfnamefont {I.}~\bibnamefont {Matsubara}}, \ and\
  \bibinfo {author} {\bibfnamefont {S.}~\bibnamefont {Sodeoka}},\ }\href
  {\doibase 10.1016/s0025-5408(00)00441-4} {\bibfield  {journal} {\bibinfo
  {journal} {Mater. Res. Bull.}\ }\textbf {\bibinfo {volume} {35}},\ \bibinfo
  {pages} {2371} (\bibinfo {year} {2000})}\BibitemShut {NoStop}%
\bibitem [{\citenamefont {Nagira}\ \emph {et~al.}(2004)\citenamefont {Nagira},
  \citenamefont {Ito},\ and\ \citenamefont {Hara}}]{Nagira2004}%
  \BibitemOpen
  \bibfield  {author} {\bibinfo {author} {\bibfnamefont {T.}~\bibnamefont
  {Nagira}}, \bibinfo {author} {\bibfnamefont {M.}~\bibnamefont {Ito}}, \ and\
  \bibinfo {author} {\bibfnamefont {S.}~\bibnamefont {Hara}},\ }\href {\doibase
  10.2320/matertrans.45.1339} {\bibfield  {journal} {\bibinfo  {journal}
  {Mater. Trans.}\ }\textbf {\bibinfo {volume} {45}},\ \bibinfo {pages} {1339}
  (\bibinfo {year} {2004})}\BibitemShut {NoStop}%
\bibitem [{\citenamefont {Seetawan}\ \emph {et~al.}(2006)\citenamefont
  {Seetawan}, \citenamefont {Amornkitbamrung}, \citenamefont {Burinprakhon},
  \citenamefont {Maensiri}, \citenamefont {Kurosaki}, \citenamefont {Muta},
  \citenamefont {Uno},\ and\ \citenamefont {Yamanaka}}]{Seetawan2006}%
  \BibitemOpen
  \bibfield  {author} {\bibinfo {author} {\bibfnamefont {T.}~\bibnamefont
  {Seetawan}}, \bibinfo {author} {\bibfnamefont {V.}~\bibnamefont
  {Amornkitbamrung}}, \bibinfo {author} {\bibfnamefont {T.}~\bibnamefont
  {Burinprakhon}}, \bibinfo {author} {\bibfnamefont {S.}~\bibnamefont
  {Maensiri}}, \bibinfo {author} {\bibfnamefont {K.}~\bibnamefont {Kurosaki}},
  \bibinfo {author} {\bibfnamefont {H.}~\bibnamefont {Muta}}, \bibinfo {author}
  {\bibfnamefont {M.}~\bibnamefont {Uno}}, \ and\ \bibinfo {author}
  {\bibfnamefont {S.}~\bibnamefont {Yamanaka}},\ }\href {\doibase
  10.1016/j.jallcom.2005.06.032} {\bibfield  {journal} {\bibinfo  {journal} {J.
  Alloy. Compd.}\ }\textbf {\bibinfo {volume} {407}},\ \bibinfo {pages} {314}
  (\bibinfo {year} {2006})}\BibitemShut {NoStop}%
\bibitem [{\citenamefont {Ito}\ \emph {et~al.}(2006)\citenamefont {Ito},
  \citenamefont {Nagira},\ and\ \citenamefont {Hara}}]{Ito2006}%
  \BibitemOpen
  \bibfield  {author} {\bibinfo {author} {\bibfnamefont {M.}~\bibnamefont
  {Ito}}, \bibinfo {author} {\bibfnamefont {T.}~\bibnamefont {Nagira}}, \ and\
  \bibinfo {author} {\bibfnamefont {S.}~\bibnamefont {Hara}},\ }\href {\doibase
  10.1016/j.jallcom.2004.12.110} {\bibfield  {journal} {\bibinfo  {journal} {J.
  Alloy. Compd.}\ }\textbf {\bibinfo {volume} {408}},\ \bibinfo {pages} {1217}
  (\bibinfo {year} {2006})}\BibitemShut {NoStop}%
\bibitem [{\citenamefont {Park}\ \emph {et~al.}(2006)\citenamefont {Park},
  \citenamefont {Jang}, \citenamefont {Kwon}, \citenamefont {Kim},\ and\
  \citenamefont {Cho}}]{Park2006}%
  \BibitemOpen
  \bibfield  {author} {\bibinfo {author} {\bibfnamefont {K.}~\bibnamefont
  {Park}}, \bibinfo {author} {\bibfnamefont {K.~U.}\ \bibnamefont {Jang}},
  \bibinfo {author} {\bibfnamefont {H.~C.}\ \bibnamefont {Kwon}}, \bibinfo
  {author} {\bibfnamefont {J.~G.}\ \bibnamefont {Kim}}, \ and\ \bibinfo
  {author} {\bibfnamefont {W.~S.}\ \bibnamefont {Cho}},\ }\href {\doibase
  10.1016/j.jallcom.2005.08.081} {\bibfield  {journal} {\bibinfo  {journal} {J.
  Alloy. Compd.}\ }\textbf {\bibinfo {volume} {419}},\ \bibinfo {pages} {213}
  (\bibinfo {year} {2006})}\BibitemShut {NoStop}%
\bibitem [{\citenamefont {Ito}\ and\ \citenamefont {Furumoto}(2008)}]{Ito2008}%
  \BibitemOpen
  \bibfield  {author} {\bibinfo {author} {\bibfnamefont {M.}~\bibnamefont
  {Ito}}\ and\ \bibinfo {author} {\bibfnamefont {D.}~\bibnamefont {Furumoto}},\
  }\href {\doibase 10.1016/j.jallcom.2006.11.032} {\bibfield  {journal}
  {\bibinfo  {journal} {J. Alloy. Compd.}\ }\textbf {\bibinfo {volume} {450}},\
  \bibinfo {pages} {494} (\bibinfo {year} {2008})}\BibitemShut {NoStop}%
\bibitem [{\citenamefont {Park}\ and\ \citenamefont {Lee}(2008)}]{Park2008}%
  \BibitemOpen
  \bibfield  {author} {\bibinfo {author} {\bibfnamefont {K.}~\bibnamefont
  {Park}}\ and\ \bibinfo {author} {\bibfnamefont {J.~H.}\ \bibnamefont {Lee}},\
  }\href {\doibase 10.1016/j.matlet.2007.11.090} {\bibfield  {journal}
  {\bibinfo  {journal} {Mater. Lett.}\ }\textbf {\bibinfo {volume} {62}},\
  \bibinfo {pages} {2366} (\bibinfo {year} {2008})}\BibitemShut {NoStop}%
\bibitem [{\citenamefont {Wang}\ \emph {et~al.}(2009)\citenamefont {Wang},
  \citenamefont {Wang},\ and\ \citenamefont {Zhao}}]{Wang2009}%
  \BibitemOpen
  \bibfield  {author} {\bibinfo {author} {\bibfnamefont {L.}~\bibnamefont
  {Wang}}, \bibinfo {author} {\bibfnamefont {M.}~\bibnamefont {Wang}}, \ and\
  \bibinfo {author} {\bibfnamefont {D.}~\bibnamefont {Zhao}},\ }\href {\doibase
  10.1016/j.jallcom.2008.04.013} {\bibfield  {journal} {\bibinfo  {journal} {J.
  Alloy. Compd.}\ }\textbf {\bibinfo {volume} {471}},\ \bibinfo {pages} {519}
  (\bibinfo {year} {2009})}\BibitemShut {NoStop}%
\bibitem [{\citenamefont {Tsai}\ \emph {et~al.}(2011)\citenamefont {Tsai},
  \citenamefont {Zhang}, \citenamefont {Donelson}, \citenamefont {Tan},\ and\
  \citenamefont {Li}}]{Tsai2011}%
  \BibitemOpen
  \bibfield  {author} {\bibinfo {author} {\bibfnamefont {P.~H.}\ \bibnamefont
  {Tsai}}, \bibinfo {author} {\bibfnamefont {T.~S.}\ \bibnamefont {Zhang}},
  \bibinfo {author} {\bibfnamefont {R.}~\bibnamefont {Donelson}}, \bibinfo
  {author} {\bibfnamefont {T.~T.}\ \bibnamefont {Tan}}, \ and\ \bibinfo
  {author} {\bibfnamefont {S.}~\bibnamefont {Li}},\ }\href {\doibase
  10.1016/j.jallcom.2011.02.045} {\bibfield  {journal} {\bibinfo  {journal} {J.
  Alloy. Compd.}\ }\textbf {\bibinfo {volume} {509}},\ \bibinfo {pages} {5183}
  (\bibinfo {year} {2011})}\BibitemShut {NoStop}%
\bibitem [{\citenamefont {Delley}(1990)}]{Delley1990}%
  \BibitemOpen
  \bibfield  {author} {\bibinfo {author} {\bibfnamefont {B.}~\bibnamefont
  {Delley}},\ }\href {\doibase 10.1063/1.458452} {\bibfield  {journal}
  {\bibinfo  {journal} {J. Chem. Phys.}\ }\textbf {\bibinfo {volume} {92}},\
  \bibinfo {pages} {508} (\bibinfo {year} {1990})}\BibitemShut {NoStop}%
\bibitem [{\citenamefont {Delley}(2000)}]{Delley2000}%
  \BibitemOpen
  \bibfield  {author} {\bibinfo {author} {\bibfnamefont {B.}~\bibnamefont
  {Delley}},\ }\href {\doibase 10.1063/1.1316015} {\bibfield  {journal}
  {\bibinfo  {journal} {J. Chem. Phys.}\ }\textbf {\bibinfo {volume} {113}},\
  \bibinfo {pages} {7756} (\bibinfo {year} {2000})}\BibitemShut {NoStop}%
\bibitem [{\citenamefont {Perdew}\ and\ \citenamefont
  {Wang}(1992)}]{Perdew1992}%
  \BibitemOpen
  \bibfield  {author} {\bibinfo {author} {\bibfnamefont {J.}~\bibnamefont
  {Perdew}}\ and\ \bibinfo {author} {\bibfnamefont {Y.}~\bibnamefont {Wang}},\
  }\href@noop {} {\bibfield  {journal} {\bibinfo  {journal} {Phys. Rev. B}\
  }\textbf {\bibinfo {volume} {45}},\ \bibinfo {pages} {13244} (\bibinfo {year}
  {1992})}\BibitemShut {NoStop}%
\bibitem [{\citenamefont {Zandbergen}\ \emph {et~al.}(2004)\citenamefont
  {Zandbergen}, \citenamefont {Foo}, \citenamefont {Xu}, \citenamefont
  {Kumar},\ and\ \citenamefont {Cava}}]{Zandbergen2004}%
  \BibitemOpen
  \bibfield  {author} {\bibinfo {author} {\bibfnamefont {H.~W.}\ \bibnamefont
  {Zandbergen}}, \bibinfo {author} {\bibfnamefont {M.}~\bibnamefont {Foo}},
  \bibinfo {author} {\bibfnamefont {Q.}~\bibnamefont {Xu}}, \bibinfo {author}
  {\bibfnamefont {V.}~\bibnamefont {Kumar}}, \ and\ \bibinfo {author}
  {\bibfnamefont {R.~J.}\ \bibnamefont {Cava}},\ }\href@noop {} {\bibfield
  {journal} {\bibinfo  {journal} {Phys. Rev. B}\ }\textbf {\bibinfo {volume}
  {70}},\ \bibinfo {pages} {024101} (\bibinfo {year} {2004})}\BibitemShut
  {NoStop}%
\bibitem [{\citenamefont {Zhang}\ \emph {et~al.}(2005)\citenamefont {Zhang},
  \citenamefont {Capaz}, \citenamefont {Cohen},\ and\ \citenamefont
  {Louie}}]{Zhang2005}%
  \BibitemOpen
  \bibfield  {author} {\bibinfo {author} {\bibfnamefont {P.}~\bibnamefont
  {Zhang}}, \bibinfo {author} {\bibfnamefont {R.~B.}\ \bibnamefont {Capaz}},
  \bibinfo {author} {\bibfnamefont {M.~L.}\ \bibnamefont {Cohen}}, \ and\
  \bibinfo {author} {\bibfnamefont {S.~G.}\ \bibnamefont {Louie}},\ }\href@noop
  {} {\bibfield  {journal} {\bibinfo  {journal} {Phys. Rev. B}\ }\textbf
  {\bibinfo {volume} {71}},\ \bibinfo {pages} {153102} (\bibinfo {year}
  {2005})}\BibitemShut {NoStop}%
\bibitem [{\citenamefont {Meng}\ \emph {et~al.}(2005)\citenamefont {Meng},
  \citenamefont {Van~der Ven}, \citenamefont {Chan},\ and\ \citenamefont
  {Ceder}}]{Meng2005}%
  \BibitemOpen
  \bibfield  {author} {\bibinfo {author} {\bibfnamefont {Y.~S.}\ \bibnamefont
  {Meng}}, \bibinfo {author} {\bibfnamefont {A.}~\bibnamefont {Van~der Ven}},
  \bibinfo {author} {\bibfnamefont {M.~K.~Y.}\ \bibnamefont {Chan}}, \ and\
  \bibinfo {author} {\bibfnamefont {G.}~\bibnamefont {Ceder}},\ }\href@noop {}
  {\bibfield  {journal} {\bibinfo  {journal} {Phys. Rev. B}\ }\textbf {\bibinfo
  {volume} {72}},\ \bibinfo {pages} {172103} (\bibinfo {year}
  {2005})}\BibitemShut {NoStop}%
\bibitem [{\citenamefont {Assadi}\ and\ \citenamefont
  {Katayama-Yoshida}(2015)}]{Assadi2015}%
  \BibitemOpen
  \bibfield  {author} {\bibinfo {author} {\bibfnamefont {M.~H.~N.}\
  \bibnamefont {Assadi}}\ and\ \bibinfo {author} {\bibfnamefont
  {H.}~\bibnamefont {Katayama-Yoshida}},\ }\href@noop {} {\bibfield  {journal}
  {\bibinfo  {journal} {Funct. Mater. Lett.}\ }\textbf {\bibinfo {volume}
  {08}},\ \bibinfo {pages} {1540016} (\bibinfo {year} {2015})}\BibitemShut
  {NoStop}%
\bibitem [{\citenamefont {Assadi}\ \emph {et~al.}(2013)\citenamefont {Assadi},
  \citenamefont {Li},\ and\ \citenamefont {Yu}}]{Assadi2013}%
  \BibitemOpen
  \bibfield  {author} {\bibinfo {author} {\bibfnamefont {M.~H.~N.}\
  \bibnamefont {Assadi}}, \bibinfo {author} {\bibfnamefont {S.}~\bibnamefont
  {Li}}, \ and\ \bibinfo {author} {\bibfnamefont {A.~B.}\ \bibnamefont {Yu}},\
  }\href {\doibase 10.1039/c2ra22514j} {\bibfield  {journal} {\bibinfo
  {journal} {RSC Adv.}\ }\textbf {\bibinfo {volume} {3}},\ \bibinfo {pages}
  {1442} (\bibinfo {year} {2013})}\BibitemShut {NoStop}%
\bibitem [{\citenamefont {Lang}\ \emph {et~al.}(2005)\citenamefont {Lang},
  \citenamefont {Bobroff}, \citenamefont {Alloul}, \citenamefont {Mendels},
  \citenamefont {Blanchard},\ and\ \citenamefont {Collin}}]{Lang2005}%
  \BibitemOpen
  \bibfield  {author} {\bibinfo {author} {\bibfnamefont {G.}~\bibnamefont
  {Lang}}, \bibinfo {author} {\bibfnamefont {J.}~\bibnamefont {Bobroff}},
  \bibinfo {author} {\bibfnamefont {H.}~\bibnamefont {Alloul}}, \bibinfo
  {author} {\bibfnamefont {P.}~\bibnamefont {Mendels}}, \bibinfo {author}
  {\bibfnamefont {N.}~\bibnamefont {Blanchard}}, \ and\ \bibinfo {author}
  {\bibfnamefont {G.}~\bibnamefont {Collin}},\ }\href@noop {} {\bibfield
  {journal} {\bibinfo  {journal} {Phys. Rev. B}\ }\textbf {\bibinfo {volume}
  {72}},\ \bibinfo {pages} {094404} (\bibinfo {year} {2005})}\BibitemShut
  {NoStop}%
\bibitem [{\citenamefont {Weller}\ \emph {et~al.}(2009)\citenamefont {Weller},
  \citenamefont {Sacchetti}, \citenamefont {Ott}, \citenamefont
  {Mattenberger},\ and\ \citenamefont {Batlogg}}]{Weller2009}%
  \BibitemOpen
  \bibfield  {author} {\bibinfo {author} {\bibfnamefont {M.}~\bibnamefont
  {Weller}}, \bibinfo {author} {\bibfnamefont {A.}~\bibnamefont {Sacchetti}},
  \bibinfo {author} {\bibfnamefont {H.~R.}\ \bibnamefont {Ott}}, \bibinfo
  {author} {\bibfnamefont {K.}~\bibnamefont {Mattenberger}}, \ and\ \bibinfo
  {author} {\bibfnamefont {B.}~\bibnamefont {Batlogg}},\ }\href {\doibase
  10.1103/PhysRevLett.102.056401} {\bibfield  {journal} {\bibinfo  {journal}
  {Phys. Rev. Lett.}\ }\textbf {\bibinfo {volume} {102}},\ \bibinfo {pages}
  {056401} (\bibinfo {year} {2009})}\BibitemShut {NoStop}%
\bibitem [{\citenamefont {Tsai}\ \emph {et~al.}(2012)\citenamefont {Tsai},
  \citenamefont {Assadi}, \citenamefont {Zhang}, \citenamefont {Ulrich},
  \citenamefont {Tan}, \citenamefont {Donelson},\ and\ \citenamefont
  {Li}}]{Tsai2012}%
  \BibitemOpen
  \bibfield  {author} {\bibinfo {author} {\bibfnamefont {P.~H.}\ \bibnamefont
  {Tsai}}, \bibinfo {author} {\bibfnamefont {M.~H.~N.}\ \bibnamefont {Assadi}},
  \bibinfo {author} {\bibfnamefont {T.}~\bibnamefont {Zhang}}, \bibinfo
  {author} {\bibfnamefont {C.}~\bibnamefont {Ulrich}}, \bibinfo {author}
  {\bibfnamefont {T.~T.}\ \bibnamefont {Tan}}, \bibinfo {author} {\bibfnamefont
  {R.}~\bibnamefont {Donelson}}, \ and\ \bibinfo {author} {\bibfnamefont
  {S.}~\bibnamefont {Li}},\ }\href {\doibase 10.1021/jp209343v} {\bibfield
  {journal} {\bibinfo  {journal} {J. Phys. Chem. C}\ }\textbf {\bibinfo
  {volume} {116}},\ \bibinfo {pages} {4324} (\bibinfo {year}
  {2012})}\BibitemShut {NoStop}%
\bibitem [{\citenamefont {Porter}\ \emph {et~al.}(2014)\citenamefont {Porter},
  \citenamefont {Roger}, \citenamefont {Gutmann}, \citenamefont {Uthayakumar},
  \citenamefont {Prabhakaran}, \citenamefont {Boothroyd}, \citenamefont
  {Pandiyan},\ and\ \citenamefont {Goff}}]{Porter2014}%
  \BibitemOpen
  \bibfield  {author} {\bibinfo {author} {\bibfnamefont {D.~G.}\ \bibnamefont
  {Porter}}, \bibinfo {author} {\bibfnamefont {M.}~\bibnamefont {Roger}},
  \bibinfo {author} {\bibfnamefont {M.~J.}\ \bibnamefont {Gutmann}}, \bibinfo
  {author} {\bibfnamefont {S.}~\bibnamefont {Uthayakumar}}, \bibinfo {author}
  {\bibfnamefont {D.}~\bibnamefont {Prabhakaran}}, \bibinfo {author}
  {\bibfnamefont {A.~T.}\ \bibnamefont {Boothroyd}}, \bibinfo {author}
  {\bibfnamefont {M.~S.}\ \bibnamefont {Pandiyan}}, \ and\ \bibinfo {author}
  {\bibfnamefont {J.~P.}\ \bibnamefont {Goff}},\ }\href {\doibase
  10.1103/PhysRevB.90.054101} {\bibfield  {journal} {\bibinfo  {journal} {Phys.
  Rev. B}\ }\textbf {\bibinfo {volume} {90}},\ \bibinfo {pages} {054101}
  (\bibinfo {year} {2014})}\BibitemShut {NoStop}%
\bibitem [{\citenamefont {Voneshen}\ \emph {et~al.}(2013)\citenamefont
  {Voneshen}, \citenamefont {Refson}, \citenamefont {Borissenko}, \citenamefont
  {Krisch}, \citenamefont {Bosak}, \citenamefont {Piovano}, \citenamefont
  {Cemal}, \citenamefont {Enderle}, \citenamefont {Gutmann}, \citenamefont
  {Hoesch}, \citenamefont {Roger}, \citenamefont {Gannon}, \citenamefont
  {Boothroyd}, \citenamefont {Uthayakumar}, \citenamefont {Porter},\ and\
  \citenamefont {Goff}}]{Voneshen2013}%
  \BibitemOpen
  \bibfield  {author} {\bibinfo {author} {\bibfnamefont {D.~J.}\ \bibnamefont
  {Voneshen}}, \bibinfo {author} {\bibfnamefont {K.}~\bibnamefont {Refson}},
  \bibinfo {author} {\bibfnamefont {E.}~\bibnamefont {Borissenko}}, \bibinfo
  {author} {\bibfnamefont {M.}~\bibnamefont {Krisch}}, \bibinfo {author}
  {\bibfnamefont {A.}~\bibnamefont {Bosak}}, \bibinfo {author} {\bibfnamefont
  {A.}~\bibnamefont {Piovano}}, \bibinfo {author} {\bibfnamefont
  {E.}~\bibnamefont {Cemal}}, \bibinfo {author} {\bibfnamefont
  {M.}~\bibnamefont {Enderle}}, \bibinfo {author} {\bibfnamefont {M.~J.}\
  \bibnamefont {Gutmann}}, \bibinfo {author} {\bibfnamefont {M.}~\bibnamefont
  {Hoesch}}, \bibinfo {author} {\bibfnamefont {M.}~\bibnamefont {Roger}},
  \bibinfo {author} {\bibfnamefont {L.}~\bibnamefont {Gannon}}, \bibinfo
  {author} {\bibfnamefont {A.~T.}\ \bibnamefont {Boothroyd}}, \bibinfo {author}
  {\bibfnamefont {S.}~\bibnamefont {Uthayakumar}}, \bibinfo {author}
  {\bibfnamefont {D.~G.}\ \bibnamefont {Porter}}, \ and\ \bibinfo {author}
  {\bibfnamefont {J.~P.}\ \bibnamefont {Goff}},\ }\href {\doibase
  10.1038/nmat3739} {\bibfield  {journal} {\bibinfo  {journal} {Nat. Mater.}\
  }\textbf {\bibinfo {volume} {12}},\ \bibinfo {pages} {1027} (\bibinfo {year}
  {2013})}\BibitemShut {NoStop}%
\bibitem [{\citenamefont {Chen}\ \emph {et~al.}(2004)\citenamefont {Chen},
  \citenamefont {Chen}, \citenamefont {Maljuk}, \citenamefont {Kulakov},
  \citenamefont {Zhang}, \citenamefont {Lemmens},\ and\ \citenamefont
  {Lin}}]{Chen2004}%
  \BibitemOpen
  \bibfield  {author} {\bibinfo {author} {\bibfnamefont {D.~P.}\ \bibnamefont
  {Chen}}, \bibinfo {author} {\bibfnamefont {H.~C.}\ \bibnamefont {Chen}},
  \bibinfo {author} {\bibfnamefont {A.}~\bibnamefont {Maljuk}}, \bibinfo
  {author} {\bibfnamefont {A.}~\bibnamefont {Kulakov}}, \bibinfo {author}
  {\bibfnamefont {H.}~\bibnamefont {Zhang}}, \bibinfo {author} {\bibfnamefont
  {P.}~\bibnamefont {Lemmens}}, \ and\ \bibinfo {author} {\bibfnamefont
  {C.~T.}\ \bibnamefont {Lin}},\ }\href {\doibase 10.1103/PhysRevB.70.024506}
  {\bibfield  {journal} {\bibinfo  {journal} {Phys. Rev. B}\ }\textbf {\bibinfo
  {volume} {70}},\ \bibinfo {pages} {024506} (\bibinfo {year}
  {2004})}\BibitemShut {NoStop}%
\bibitem [{\citenamefont {Mo}\ \emph {et~al.}(2014)\citenamefont {Mo},
  \citenamefont {Ong},\ and\ \citenamefont {Ceder}}]{Mo2014}%
  \BibitemOpen
  \bibfield  {author} {\bibinfo {author} {\bibfnamefont {Y.}~\bibnamefont
  {Mo}}, \bibinfo {author} {\bibfnamefont {S.~P.}\ \bibnamefont {Ong}}, \ and\
  \bibinfo {author} {\bibfnamefont {G.}~\bibnamefont {Ceder}},\ }\href
  {\doibase 10.1021/cm501563f} {\bibfield  {journal} {\bibinfo  {journal}
  {Chem. Mater.}\ }\textbf {\bibinfo {volume} {26}},\ \bibinfo {pages} {5208}
  (\bibinfo {year} {2014})}\BibitemShut {NoStop}%
\bibitem [{\citenamefont {Han}\ \emph {et~al.}(2015)\citenamefont {Han},
  \citenamefont {Lim}, \citenamefont {Jeong}, \citenamefont {Ahn},
  \citenamefont {Park}, \citenamefont {Sohn},\ and\ \citenamefont
  {Pyo}}]{Han2015}%
  \BibitemOpen
  \bibfield  {author} {\bibinfo {author} {\bibfnamefont {S.~C.}\ \bibnamefont
  {Han}}, \bibinfo {author} {\bibfnamefont {H.}~\bibnamefont {Lim}}, \bibinfo
  {author} {\bibfnamefont {J.}~\bibnamefont {Jeong}}, \bibinfo {author}
  {\bibfnamefont {D.}~\bibnamefont {Ahn}}, \bibinfo {author} {\bibfnamefont
  {W.~B.}\ \bibnamefont {Park}}, \bibinfo {author} {\bibfnamefont {K.-S.}\
  \bibnamefont {Sohn}}, \ and\ \bibinfo {author} {\bibfnamefont
  {M.}~\bibnamefont {Pyo}},\ }\href {\doibase
  http://dx.doi.org/10.1016/j.jpowsour.2014.11.150} {\bibfield  {journal}
  {\bibinfo  {journal} {J. Power Sources}\ }\textbf {\bibinfo {volume} {277}},\
  \bibinfo {pages} {9} (\bibinfo {year} {2015})}\BibitemShut {NoStop}%
\end{thebibliography}%

\end{document}